\documentclass[pre,preprint,english]{revtex4}
\usepackage{fancyhdr}
\usepackage{amsmath,amsfonts,amssymb}
\usepackage{graphicx}
\usepackage{color}
\usepackage{bbm}

\begin{document}
\title{Towards a semiclassical understanding of chaotic single- and many-particle quantum 
          dynamics at post-Heisenberg time scales}
\author{Daniel Waltner}
\address{Fakult\"at f\"ur Physik, Universit\"at Duisburg-Essen, Lotharstra\ss e 1, 47048 Duisburg, Germany}
\author{Klaus Richter}
\address{Institut f\"ur Theoretische Physik, Universit\"at Regensburg, D-93040 Regensburg, Germany}


\begin{abstract}
Despite considerable progress during the last decades in devising a semiclassical theory for classically 
chaotic quantum systems a quantitative semiclassical understanding of their dynamics at late times 
(beyond the so-called Heisenberg time $T_H$) is still missing. 
This challenge, corresponding to resolving spectral structures on energy scales below the mean level spacing, 
is intimately related to the quest for semiclassically restoring quantum unitarity, which is
reflected in real-valued spectral determinants. 
Guided through insights for quantum graphs we devise a periodic-orbit resummation procedure for 
quantum chaotic systems invoking periodic-orbit self encounters as the structuring element of a hierarchical 
phase space dynamics.
We propose a way to purely semiclassically construct real spectral determinants 
based on two major underlying mechanisms: (i) Complementary contributions to the spectral determinant 
from regrouped pseudo orbits of duration $T < T_H$ and $T_H-T$ are complex conjugate to each other.
(ii) Contributions from long periodic orbits involving multiple traversals along shorter orbits cancel out. 
We furthermore discuss implications for interacting $N$-particle quantum systems with a chaotic 
classical large-$N$ limit that have recently attracted interest in the context of many-body quantum chaos.
\end{abstract}
\maketitle

\section{Introduction}
The close connection between the energy level spectra of quantum systems with a chaotic classical limit
and those resulting from random matrix theory was originally found by Bohigas, Giannoni and Schmit \cite{Boh84} 
who conjectured that the level statistics of classically chaotic quantum systems obeys universal 
random matrix predictions. 
One major approach to justify this conjecture is based on periodic orbit (PO) expansions of the spectral density 
(in terms of the Gutzwiller trace formula \cite{Gut90}) or related quantities such as the spectral determinant 
or zeta function \cite{BeK,Bogomo,Kea88,Cvita,Wirzba,Cvitano}.
For certain systems such expansions are exact, for example for the dynamics on surfaces of constant negative 
curvature \cite{Bal86}. Generally, such PO expansions represent appropriate approximations 
to quantum spectra in the semiclassical limit that is defined by the condition that actions 
(of the shortest POs) in the corresponding classical system are much larger than the action 
quantum $\hbar$. 

Furthermore, correlation functions of the spectral density or spectral determinant, both as functions
of the energy $E$, can then be expressed in terms of multiple sums over contributions from the POs of 
the considered system. 
Each of the summands involved carry phases depending on the difference of the actions of different
POs divided by $\hbar$. 
As long as the POs involved are classically uncorrelated this results in phases rapidly varying as a function 
of $E$ such that the corresponding joined PO contributions average out. 
This lead to the quest for identifying pairs of correlated POs with non-random phase differences that survive
the energy averages involved. This program was initiated by Berry \cite{Ber85} showing by means of the 
Hannay-Ozorio de Almeida sum rule \cite{Han84} that the leading-order contribution to the spectral 
form factor, {\em i.e.}\ the Fourier transform of the spectral two-point  auto-correlation function,
is obtained by pairing identical trajectories, known as the so-called diagonal approximation.
This approach is adequate for POs with periods $T$ considerably smaller than the Heisenberg time $T_H$. 
This corresponds to the time dual to the mean energy level spacing 
$\delta(E)$, {\em i.e.} the inverse of the mean level density $\overline{\rho}(E)$.




Further semiclassical contributions to the form factor for time scales $\tau=T/T_H<1$ result from pairs of POs linked
to each other via close self encounters where the periodic trajectories approach and depart from each 
other exponentially fast in the corresponding so-called encounter regions.
The structure of the underlying trajectories was identified in Ref.~\cite{Sie01} for the leading-order 
off-diagonal contribution to the spectral form factor.
All off-diagonal higher-order corrections 
were eventually treated in \cite{Mul04} leading to the complete result for the spectral 
form factor for $\tau<1$, in accordance with random matrix theory. The latter work also established 
an analogy between PO diagrams and diagrams occurring in a perturbative expansion of the corresponding
field theory. This concept found many applications in mesoscopic physics, see e.g.\ \cite{Kuipersm,Brouwerm,Essenm,Jacquodm,Novaes}. 

The remaining challenge has been to investigate the regime of times $T>T_H$,
corresponding to semiclassically resolving structures beyond the mean level spacing. 
The form factor for $\tau>1$ was then formally obtained in a way that was inspired by a diagrammatic expansion of 
a four-point correlation function of spectral determinants around a second saddle point of the corresponding 
field theory \cite{Heu07}. Later on this procedure was connected to the use of the Riemann Siegel lookalike 
formula for the spectral determinant \cite{Kea07} and to analytic continuation \cite{BrWa}. 
All such approaches have in common that they extend semiclassical results for $\tau<1$ to the regime
$\tau>1$ by additionally invoking quantum mechanical unitarity in one or the other way. 
To the best of our knowledge a dynamical justification, {\em i.e.}\ a derivation entirely 
based on a semiclassical theory in terms of correlated POs, is however still lacking. 
This is much more than an academic problem and involves the deep and still open, far-reaching question in 
how far quantum unitarity can be achieved in an entirely semiclassical way, at least
 for chaotic quantum systems. This question is also particularly relevant 
for semiclassical calculations of the spectra of quantum chaotic systems themselves and not only
for spectral correlations. Implications of unitarity were studied in this context in Refs.\
\cite{Marden,Bog,Kus,Schomerus,Wal13}, and there exist also close connections to number theory \cite{KeaBer}.
Moreover, in the context of black hole physics the saturation and late-time behavior of the spectral form 
factor has recently attracted interest in theories with a graivtational dual motivated by the black hole 
information problem \cite{Stanford}.

\begin{figure}
 \begin{center}
 \includegraphics[width=6cm]{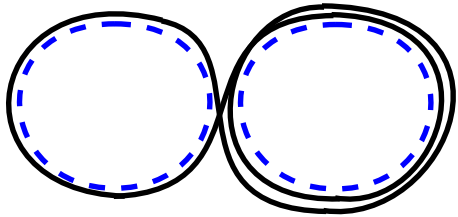}
\caption{Partial repetitions of a periodic orbit: A longer orbit (solid curve) comprising a 2-encounter formed 
by two shorter periodic orbits (dashed lines) traverses one of the shorter periodic orbits twice (dashed).}
\label{fig3}
\end{center}
\end{figure}

The developments mentioned above have lead to the believe that there is a complete analogy between 
field theory on the one side and semiclassical PO expansions on the other side.
However, there are also important differences between the two methods: The effect of integrating over 
a curved manifold in field theory found its correspondence in the PO approach as contributions from long
POs partially multiply following other shorter POs \cite{Wal09} (see {\em e.g.} Fig.~\ref{fig3})
referred to as partial repetitions in the following. Such configurations 
turn out in Refs.\ \cite{Mul04,Heu07} to eventually cancel out for the form factor for $\tau < 1$.
We will discuss them in detail below.

Quantum graphs \cite{Kottos} represent another class of systems where semiclassical approaches turn out
to be exact and silent features of unitarity can be conveniently explored.
Quantum graphs consist of a network of bonds along which the dynamics is usually determined by the one-dimensional 
Schr\"odinger equation. Different bonds are inter-connected through vertices with 
couplings modelled by a unitary scattering matrix. 
Graphs have proven very useful as simple model systems in quantum chaos. For these systems a resummation 
procedure for the spectral determinant could be achieved that is purely based on the dynamics of the 
system \cite{Wal13}. As a key element it involves POs following shorter POs several times.
It can thus be anticipated that contributions from different saddle points in field theory find their 
correspondence as orbits containing higher repetitions of shorter orbits in semiclassical PO
expansions (as depicted in Fig.\ \ref{fig3}). For quantum graphs such configurations were shown to provide the
underlying mechanism for the saturation of the form factor beyond $T_H$ \cite{Wal13}.


However, as far as we know, such a correspondence has not yet been established for other 
chaotic systems, e.g.\ billiards, maps and surfaces of constant negative curvature. 
Carrying over the procedure for graphs to the latter systems is not at all straightforward 
as the phases from the unitary scattering matrices at the vertices 
entering the spectral determinants of quantum graphs find no direct correspondence there.

The aim of this paper is to fill this gap:
We suggest a corresponding resummation procedure for 
general chaotic systems. To be more precise, in our derivation we require ergodicity and hyperbolicity 
of the underlying classical dynamics such that a statistical description of close self encounters is possible, 
for more details on the requirements, see Ref.~\cite{Mul04}. Within our approach we {\it do not assume} 
but {\it provide} a Riemann Siegel look-alike relation for the spectral determinant indicating
the validity of the Bohigas Giannoni Schmit conjecture \cite{Boh84} based purely on PO
correlations without additionally invoking quantum mechanical unitarity.
 
The paper is organized as follows: In the next section we define the relevant quantities,
explain what we mean by resummation and recapitulate basic steps of the resummation procedure for quantum graphs. 
In Sec.~\ref{resumgen} we show how to perform resummation for general chaotic systems. 
In Sec.~\ref{MB} we discuss implications for classically chaotic many-body quantum systems.
We conclude in Sec.~\ref{conclu} followed by an appendix containing technical details.

\section{General Aspects of Resummation}
\subsection{Spectral Determinant}
\label{subsec22}

The spectral determinant is defined as (see e.g.\ Ref.~\cite{Kea88})
\begin{equation}
\label{eq0}
 \Delta(E)=\det\left[A(E,\hat{H})\left(E-\hat{H}\right)\right]=\prod_{n=1}^\infty A(E,E_n)(E-E_n)
\end{equation}
with an appropriately chosen real regularizing function $A(E,E_n)$ that makes the product convergent.
The zeros of the spectral determinant provide the energy eigenvalues $E_n$ of the corresponding
quantum system which explains its central role.
Various periodic orbit (PO) expansions of the spectral determinant have been considered (see e.g.\ 
Refs.~\cite{BeK,Bogomo,Kea88,Cvita,Wirzba,Cvitano}). Within these approaches the spectral determinant
is usually expressed as infinite sum over classical pseudo orbits $\Gamma$ (composite POs)
of the underlying system \cite{Kea88}:
\begin{equation}
\label{eq1}
\Delta(E)={\rm e}^{-i\pi\overline{N}(E)}\sum_\Gamma (-1)^{n_\Gamma} A_\Gamma{\rm e}^{iS_\Gamma(E)/\hbar}.
\end{equation}
Here $\overline{N}(E)=\int_{-\infty}^EdE'\overline{\rho}(E')$ is the average level counting function 
with $\overline{\rho}(E')$ the mean level density and $n_\Gamma$ denotes the number of composites of the 
pseudo orbit $\Gamma$. Furthermore, $A_\Gamma$ is the product of the stability amplitudes of the orbits 
forming $\Gamma$ and $S_\Gamma(E)$ is the sum of the corresponding classical actions. We
do not consider the contributions of real repetitions (multiple traversals of the entire POs) 
to Eq.\ (\ref{eq1}) as they give subleading contributions \cite{Kea88}.
Instead longer POs partially following shorter orbits several times as shown in Fig.\ \ref{fig3} 
will prove relevant.

Expression (\ref{eq1}) is not suitable for a numerical computation of the spectrum as it contains an 
infinite sum over POs. As their number increases exponentially with length, in analogy to the Riemann-Siegel 
resummation for the corresponding formula for the zeros of the Riemann zeta function \cite{KeaBer}, an expression containing a 
truncated sum was considered based on the fact that the expression (\ref{eq0}) is real:
\begin{equation}
\label{eq10}
\Delta_{\rm res}(E)={\rm e}^{-i\pi\overline{N}(E)}\sum_{\Gamma,T\leq T_H/2} (-1)^{n_\Gamma} A_\Gamma{\rm e}^{iS_\Gamma(E)/\hbar}+c.c. \, .
\end{equation}
The sum in Eq.\ (\ref{eq10}) contains pseudo orbits only up to half of the Heisenberg time that can be expressed 
in terms of the inverse mean level spacing as $T_H=2\pi\hbar\overline{\rho}(E)$.

 In the following
we will refer to the procedure of transforming Eq.\ (\ref{eq1}) into Eq.\ (\ref{eq10}) as resummation. 
Without assuming that $\Delta(E)$ is real, that is without imposing unitarity, we will argue how
to resum and add up the contributions from pseudo orbits in a semiclassical framework in order to 
achieve $\Delta(E)$ to be real. Correspondingly, to explicitly show Eq.~(\ref{eq10}), consists of 
the following two essential steps that serve as a guideline for Sec.~\ref{resumgen}:

\begin{enumerate}
\item To demonstrate that the pseudo-orbit contributions to Eq.\ (\ref{eq1}) with durations $T_H/2<T_\Gamma\leq T_H$ provide a
contribution to Eq.\ (\ref{eq1}) complex conjugated to the one stemming from (complementary) pseudo orbits with durations 
$0\leq T_\Gamma<T_H/2$.
\item To show that all contributions to Eq.\ (\ref{eq1}) from pseudo orbits partially multiply traversing others mutually cancel.
\end{enumerate}

\subsection{Resummation for quantum graphs}\label{graphs}

As stated, Eq.\ (\ref{eq10}) can be derived from Eq.\ (\ref{eq1}) in a similar way as for the zeros of the Riemann zeta function 
(Riemann-Siegel formula) using the fact that Eq.\ (\ref{eq0}) is real. However such a formal procedure does 
not provide any insight from a semiclassical perspective into the underlying PO structures,
{\em i.e.}\ how the dynamics of the system and correlations between POs imply Eq.\ (\ref{eq10}).

\begin{figure}
\begin{center}
  \includegraphics[width=11cm]{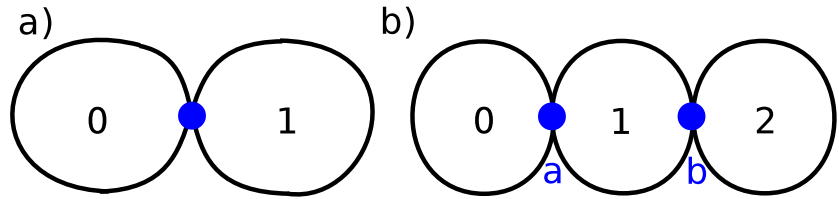}
  \end{center}
  \caption{Two simple quantum graph structures: In a) a directed graph connects 2 bonds ``0'' and ``1'' at 
one vertex. In b) the directed quantum graph consists of two vertices $a$ and $b$ connected by 4 bonds.}
\label{fig1}
\end{figure}

Such a connection was made for quantum graphs in \cite{Wal13,Band} extending earlier thorough
studies for simple graphs in \cite{Schanz,Tanner} (as the one shown in Fig.\ \ref{fig1}a)).
For quantum graphs the following resummation scheme arose:
 All contributions to the spectral determinant in Eq.\ (\ref{eq1}) from pseudo orbits involving multiple traversals 
of shorter orbits (similar to Fig.\ \ref{fig3}) cancel. This cancellation mechanism is related to the one found
by Cvitanovi\'{c} in the framework of the cycle expansion \cite{Cvitano}, but goes beyond that. 
This implies in particular for graphs, where the total bond length defines $T_H$, that POs
with duration $T>T_H$ do not contribute, since they must necessarily contain partial repetitions. 
Moreover, contributions to the spectral determinant from pseudo orbits without 
partial repetitions are related to each other in a specific way: The contributions of pseudo orbits covering
a certain part of the graph are complex conjugated to the contributions from pseudo orbits covering the complement 
(the part of the graph not covered by the pseudo orbits mentioned first).


For an illustration of this mechanism consider the directed quantum graph in Fig.\ \ref{fig1}b) containing 
the POs ``0'', ``1'', ``2'' connected by the two vertices $a$ and $b$. 
In this case $A_\Gamma$ in Eqs.\ (\ref{eq1}) and (\ref{eq10}) is the product 
of the scattering matrix elements of the vertices traversed by $\Gamma$ \cite{Wal13}. 
We first address item 1 at the end of Subsec.~\ref{subsec22}. 
To understand how complementary pseudo orbit contributions add up we consider the specific
unitary matrix 
\begin{equation}
\label{eqS}
    \mathcal{C}=\frac{1}{\sqrt{2}}\left(\begin{array}{cc}1&1\\-1&1\end{array}\right)
\end{equation}
defining the scattering at the vertex,
as for this case we will find a direct correspondence between quantum graphs and general chaotic dynamics, see below.
 
The choice of $\mathcal{C}$ in Eq.\ (\ref{eqS}) implies the following simple rule: 
Each time the (longer) orbit switches at a vertex between two (shorter) POs on the 
graph (for example from ``0'' to ``1'' in 
the graph of Fig.\ \ref{fig1}b) and back), its contribution to the spectral determinant involves the off-diagonal 
elements of $\mathcal{C}$ and acquires a minus sign, whereas a plus sign is acquired if there is no switching
between POs. 
This implies that the contributions from orbit ``01'' and from the pseudo orbit consisting of
``0'' and ``1'' {\em add up} (the minus sign from $(-1)^{n_\Gamma}$ in Eq.\ (\ref{eq1}) cancels with the one from 
the scattering matrix element). 
The stability amplitudes of the sum of the contributions from ``01'' and ``0'',``1''
are thus equal to the amplitude of the complementary orbit ``2''. The same holds true for periodic pseudo orbits traversing 4 bonds. Here the contributions 
from the pseudo orbits "012", "0" "12", "2" "01" and "0" "1" "2" add up cancelling the factor $\sqrt{2}^{-4}$
from the stability amplitudes. 
Taking into account that for the graph
\begin{equation}\label{nav}
 \pi\overline{N}(E)=\pi E\overline{\rho}(E)=\left(S_0+S_1+S_2\right)/(2\hbar),
\end{equation}
the contributions to Eq.\ (\ref{eq1}) from first ``01'',``1'',``0'' and second ``2'' on the one hand 
and first "012", "01" "2", "0" "12", "0" "1" "2" and second the pseudo orbit containing no orbit on the other hand  
altogether add up to a real quantity. 

A specific example for an orbit with multiple traversals is ``001''. Its contribution to Eq.\ (\ref{eq1}) is 
canceled by the one from the pseudo orbit consisting of ``0'' and of ``01'', as both involve the traversals 
of the same vertices and bonds differing only by the factor $(-1)^{n_\Gamma}$. 
This illustrates item 2 pointed out at the end of the last subsection for this particular  system.


\section{Resummation for general chaotic systems}\label{resumgen}

\subsection{Outline}

We now generalize the procedure just outlined to chaotic dynamical systems.
We first consider the statistics of 2-encounters for a representative PO (not a pseudo orbit) 
and will then introduce a resummation procedure along the lines of resummation outlined for quantum graphs.

The number $n$ of encounters for an orbit with $T=T_H$ can be estimated as follows:
In general chaotic systems the stability amplitude $A_\Gamma$ introduced in Eq.\ (\ref{eq1})  is given for large $T$ asymptotically by
${\rm e}^{-\lambda T_H/2}$. The number of pseudo orbits without partial repetitions can be estimated by 
$2^n$ -- at every encounter of two POs as shown in Fig.\ \ref{fig2} the longer orbit can decide to switch
to another one or stay close to the first one. From these considerations we obtain the condition

\begin{equation}
\label{eq11}
{\rm e}^{\lambda T_H/2}\approx 2^n
\end{equation}
for general chaotic systems, implying $n\approx\lambda T_H/(2\ln 2)$ or

\begin{equation}\label{napprox}
n\approx d/\hbar
\end{equation}
with a classical constant $d$. Note that this estimate for $n$ is consistent with the one that follows from 
ergodicity arguments given in \cite{Sie01}, if PO pairs with action differences up to order $\hbar$ are taken 
into account, for details see the appendix. 

In the following subsections we explain how to perform resummation, corresponding to the two steps at the end of 
Subsec.~\ref{subsec22}, for general chaotic systems.
Our reasoning is based on the statistical description of close encounters of POs
developed in Refs.\ \cite{Mul04,Heu07,Wal09}. 
To make the presentation more transparent, we split it into three parts according to the following Subsections.

In  Subsec.~\ref{en1} we ignore the relation (\ref{napprox}) for the moment and restrict ourselves to the case that
POs of duration $T_H$ possess at most one encounter. Hence
diagrams with more than one encounter are not considered here and will be treated later.
We demonstrate how in this case resummation can be performed in a similar way as in the case of a quantum graph 
containing a single vertex. More precisely, there we show how the contributions to the spectral determinant add up for durations 
equal to the Heisenberg time (contained in case 1 in Sec.~\ref{subsec22})
and how contributions from pseudo orbits with partial repetitions cancel (case 2). 

In Subsec~\ref{2en} we then consider orbits with duration $T_H$ possessing at most two encounters. 
Here we again establish the analogy to resummation for graphs containing two vertices. We  show how the contributions add up 
at durations $T=T_H$ and additionally in the range $T_H/2<T<T_H$ (case 1).
 Case 2 can be shown in the same way as in Subsec.~\ref{en1}.

In the last subsection we eventually work out the analogy between resummation for general chaotic systems 
and for quantum graphs for arbitrary $n$. This also includes $n$-values consistent with Eq.\ (\ref{napprox})
using the results from the previous subsections.

\subsection{Orbits with at most one encounter}
\label{en1}

\begin{figure}
\begin{center}
  \includegraphics[width=6cm]{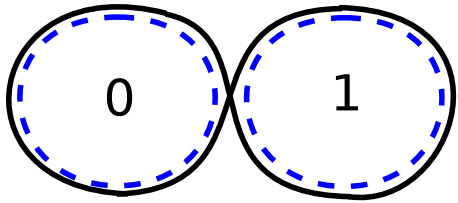}
 \caption{Sketch of a periodic orbit possessing one 2-encounter as considered for resummation in Sec.~\ref{resumgen}.}
\label{fig2}
 \end{center}
\end{figure}

We first consider orbits (as the two dashed ones in Fig.\ \ref{fig2}) possessing at most one close encounter
 in phase space, {\em i.e.\ } $n\leq1$. As established in the literature \cite{Mul04,Heu07}
the distance between the orbit segments in phase space close to the encounter is measured along the 
directions of the stable and unstable manifold of one of the shorter orbits shown as dashed lines 
in Fig.\ \ref{fig2}. The distances are denoted by $s$ and $u$ for the stable and unstable direction, respectively.

As outlined at the end of Subsec.~\ref{subsec22}, for the resummation we need to distinguish 
two scenarios that require different treatment. The first possibility contains contributions that add up from 
different pseudo orbits, based on Fig.\ \ref{fig2}. 
The second one involves orbits with higher partial repetitions along at least one of the underlying
short orbits indicated by dashed lines, an example is plotted in Fig.\ \ref{fig3}.

We start with the first scenario considering the pseudo orbit pair in Fig.\ \ref{fig2}.
The long ''eight-shaped'' PO "01", is assumed to possess the period $T_{01}=T_H$
as we consider it as the reference filling the complete phase space without any partial repetitions. 
The location of its 2-encounter divides the orbit "01" into two pieces of durations $T_0$ and $T_1$ 
of roughly the same order with $T_0+T_1\approx T_{01}$. We further need to calculate the overall contribution
to the spectral determinant (\ref{eq1}) of these two partner trajectories.
To this end we first summarize how the small difference between the pseudo orbit, composed of ''0''
and ''1'', and the long orbit is calculated on average in a statistical way.
It is quantified in terms of the distances along the stable and unstable directions $s$ and $u$ in a Poincar\'e surface of section placed within the
region where the orbits ``0'' and ``1'' encounter each other.
Here the action difference $\Delta S=su$ \cite{Mul04,Heu07} as well as the weight function $w_{01}$ determining 
the probability for such an encounter to occur in a single PO is required. 
In this context the duration of the encounter between the two short orbits
``0'' and ``1'', 
\begin{equation}\label{eq:tenc}
t_{\rm enc}=\frac{1}{\lambda}\ln\left(\frac{c^2}{|su|}\right)
\end{equation}
is relevant. It corresponds to the Ehrenfest time for $\Delta S \simeq \hbar$.
In Eq.~(\ref{eq:tenc}), $\lambda$ is the Lyapunov exponent  and $c$ a classical constant 
that determines the upper limit for the coordinates $s$ and $u$ allowing for linearization of the 
dynamics along ``0'' around the one on ``1''. The parts of the orbits ``0'' and ``1'' inside this linearizable region
are commonly referred to as encounter stretches. Then the weight function 
is given by \cite{Mul04}
\begin{equation}\label{weight}
 w_{01}(s,u)=\frac{T_{01}(T_{01}-2t_{\rm enc})}{2\Omega t_{\rm enc}}
\end{equation}
with $\Omega$ the volume of the surface of constant energy of the underlying system. 
This allows us to rewrite the common contribution from the orbit "01" and the pseudo orbit consisting of "0" and "1" 
to the spectral determinant in Eq.\ (\ref{eq1}) as
\begin{eqnarray}\label{overcon}
&&A_{{01}}{\rm e}^{iS_{{01}}/\hbar}-A_{0}A_{1}{\rm e}^{i\left(S_{0}+S_{1}\right)/\hbar}
=A_{{0}}A_{1}{\rm e}^{i\left(S_{{0}}+S_{1}\right)/\hbar}\left({\rm e}^{i\Delta S/\hbar}-1\right)
\end{eqnarray}
with $\Delta S=S_{01}-S_0-S_1$. 
Note that the orbits ``0'' and ``1'' are two representatives of the pseudo orbits $\Gamma$ in Eq.\ (\ref{eq1}) 
that we sum over when computing the spectral determinant. 
We consider the sum in Eq.~(\ref{eq1}) to act as an average over many such pairs (this will become more clear below).
This allows for replacing in Eq.~(\ref{overcon}) the phase containing the action difference by its average using 
the weight function (\ref{weight}) and $\Delta S=su$. Then we obtain
\begin{eqnarray}\label{overcon1}
 &&A_{0}A_{1}{\rm e}^{i\left(S_{0}+S_{1}\right)/\hbar}\left(\int_{-c}^cdsdu\,w_{T_{01}}(s,u){\rm e}^{i\Delta S/\hbar}-1\right)\nonumber\\&&\approx -A_{0}A_{1}{\rm e}^{i\left(S_{0}+S_{1}\right)/\hbar}
\left(\frac{T_{01}}{T_H}+1\right).
\end{eqnarray}
In the last step we employed the relation
\begin{equation}
\label{suint}
 \int_{-c}^cdsdu\,w_{01}(s,u){\rm e}^{isu/\hbar}=-\frac{T_{01}}{T_H}
\end{equation}
obtained after averaging with respect to a classically small energy window (or equivalently duration window $\Delta T_{01}$) \cite{Mul04}
leading to the fact that effectively only action differences of the order $\hbar$ contribute.
Note that only the second term in the bracket in Eq.\ (\ref{weight}) contributes to Eq.~(\ref{suint}) and
the relation $T_H=\Omega/(2\pi\hbar)$ was used.

It is instructive to compare these results with the results in Ref.~\cite{Wal13} for the quantum
graph in Fig.\ \ref{fig1}a). Denoting as in Ref.~\cite{Wal13} a contribution to the sum in Eq.~(\ref{eq1}) 
by
\begin{equation}
t_\Gamma =(-1)^{n_\Gamma}A_\Gamma {\rm e}^{iS_\Gamma/\hbar}
\end{equation}
we have the contributions from orbits in the graph that do {\em not} involve repetitions, {\em i.e.\ }
the POs $t_0$, $t_1$, $t_{01}$ and the pseudo orbit $t_0t_1$. 
In order to relate in this case Eq.\ (\ref{eq1}) to Eq.\ (\ref{eq10}) we need to consider $t_{01}$ and $t_0t_1$ 
using Eq.\ (\ref{eqS}):
\begin{equation}
 t_{01}+t_0t_1=-\left(\frac{1}{2}+\frac{1}{2}\right){\rm e}^{i\left(S_0(E)+S_1(E)\right)}=-{\rm e}^{i\left(S_0(E)+S_1(E)\right)}
\end{equation}
with $T_0+T_1=T_H$.  This contribution yields, together with the one from its complement, the pseudo orbit of length zero, a real 
contribution to Eq.\ (\ref{eq1}) when taking into account the corresponding relation (\ref{nav}) in this case. 
Note that the factor $(-1)^{n_\Gamma}$ in Eq.\ (\ref{eq1}) is canceled in graphs by the factor $(-1)$ resulting 
from the different scattering phases in Eq.~(\ref{eqS}).

Most notably, the same structure of signs is obtained in Eq.\ (\ref{overcon1}) arising from the minus sign
of the encounter integral (\ref{suint}). This integral represents an average over the many pairs of orbits 
with durations $T_0$ and $T_1$ in a chaotic dynamical system. The minus sign occuring from the encounter integral 
takes the same distinct role as the minus sign in the scattering matrix (\ref{eqS}) in quantum graphs 
required for unitarity. Hence one can regard the minus sign arising from the encounter integral to be at the
heart of a subtle semiclassical mechanism to acheive unitary behavior.
To conclude this part, in a chaotic system the semiclassical contributions to the spectral determinant
from the connected orbits add up with those from the disconnected ones in configurations shown in Fig.~\ref{fig2}
yielding the common term (\ref{overcon1}).

In the next step we show that each orbit $p$ of duration $T_p$ longer than $T_H$ indeed has at least one encounter involving 
a multiple traversal of a shorter orbit. This follows from a closer inspection of the case where one of the short orbits 
is multiply traversed as in Fig.\ \ref{fig3}. The corresponding weight function $v_{T_p}(s,u)$ for that event can
be obtained by similar arguments as in Ref.~\cite{Mul04}, see also Ref.\ \cite{Wal09}: There for an ergodic system
the number of piercing points through the Poincar\'e surface of section in the encounter region 
in the intervals $(s,s+ds)$ and $(u,u+du)$ of stable and unstable coordinates in a time
interval $(t,t+dt)$ is shown to be 
\begin{equation}\label{eq000}
 \frac{1}{\Omega}ds\,du\,dt \, .
\end{equation}
The weight function $v_{T_p}(s,u)$ is then obtained by integrating this expression over all allowed values 
of $t$ and multiplying it by $T_p/t_{\rm enc}$ in order to take into account all possible positions of the 
first encounter
 stretch.
For $v_{T_p}(s,u)$ the variable $t$ in Eq.\ (\ref{eq000}) has to be integrated from $0$ to $t_{\rm enc}$ 
as the encounter stretches are required to overlap, {\em i.e.}\ we claim that one of the two short
orbits in Fig.\ \ref{fig3} is multiply traversed by the long orbit with $T_p>T_H$ yielding
\begin{equation}\label{weight10}
 v_{T_p}(s,u)=\frac{T_pt_{\rm enc}}{t_{\rm enc}\Omega}.
\end{equation}
This weight function allows us to evaluate the sum
\begin{equation}
\label{susi}
 \sum_i{\rm e}^{is_iu_i/\hbar}
\end{equation}
running over all encounters of an orbit involving multiple traversals with action 
difference $\Delta S_i=s_iu_i$ up to order $\hbar$ \cite{Kuipers,Brouwerm}.
Replacing this sum by an integral over $s$ and $u$, and employing Eq.~(\ref{weight10}) we get
\begin{equation}
\label{siint}
 \int_{-c}^cdsdu\,v_{T_p}(s,u){\rm e}^{isu/\hbar}=\int_0^cds\frac{4\hbar 
T_p}{\Omega s}\sin\frac{sc}{\hbar}=\frac{T_p}{T_H},
\end{equation}
where we used the semiclassical approximation $c^2/\hbar\gg1$ in the last step.
Note that, in contrast to Eq.\ (\ref{suint}), in Eq.\ (\ref{siint}) no average with respect to $\Delta T_p$ 
is needed for calculating the integral.
The expression in Eq.~(\ref{siint}) follows from Eq.~(\ref{susi}) by an average using 
the distribution function $v_{T_p}(s,u)$.
According to the right hand side of (\ref{siint}) the absolute value of expression (\ref{susi}) yields a number larger than
one for $T_p>T_H$, implying that each orbit has on average at least one encounter involving a partial repetition.
Then no contributions to Eqs.\ (\ref{eq1},\ref{eq10}) arise for $T_p>T_H$ as we will show in the following.

Turning to case 2 defined at the end of Subsec.\ 2.1
we are concerned with the situation where at least one of the orbits ``0'' and ``1'' is multiply traversed. 
Then in the case of graphs the connection between Eq.\ (\ref{eq1}) and Eq.\ (\ref{eq10}) 
results from the cancellation of the contribution from a given orbit (with multiple partial traversals)
by the contribution from a corresponding pseudo orbit where one of the multiply traversed sub-orbits 
is split off \cite{Wal13}. For instance, the orbit ``001'' is canceled by the pseudo orbit
consisting of the orbits ``0'' and ``01''. In such cases both components, here ``001'' as well as 
``0'' and ``01'', exhibit the same number of transitions between ``0'' and ``1''. 

Notably, the same cancellation holds true for general chaotic dynamics: 
To show this we split the long POs in Figs.\ \ref{fig3} and \ref{fig2} into their components ``0'' and ``1'' 
getting
\begin{eqnarray}
&&A_{{001}}{\rm e}^{iS_{{001}}/\hbar}-A_{{01}}A_{0}{\rm e}^{i\left(S_{{01}}+S_{0}\right)/\hbar}\approx
A_{0}^2A_{1}{\rm e}^{i\left(2S_{0}+S_{1}\right)/\hbar}\left({\rm e}^{i\Delta S_1/\hbar}-{\rm e}^{i\Delta S_2/\hbar}\right) \, . \nonumber
\end{eqnarray}
Here $\Delta S_1$ is the action difference between the PO  ``001'' and the pseudo orbit consisting
of ``0'', ``0'' and ``1'' and, correspondingly, $\Delta S_2$ is the action difference between the 
pseudo orbits 
 ``01'' and ``0'' and again ``0'', ``0'' and ``1''. The action differences  
are given by \cite{Brouwerm,Kuipers}
\begin{equation}
 \Delta S_1=su+su{\rm e}^{-\lambda T_0},\hspace*{1cm}\Delta S_2=su \, .
\end{equation}
The terms $su$ in $\Delta S_1$ and $\Delta S_2$ are determined only by the difference in the number of 
respective transitions between ``0'' and ``1'' of the relevant pseudo orbits and are thus the same for 
$\Delta S_1$
 and $\Delta S_2$. 
Only the second term in $\Delta S_1$ depends on the details of how the corresponding pseudo orbit 
surrounds ``0'' and ``1''.
Following the same reasoning as  above Eq.~(\ref{overcon1}) we again replace the latter expression 
by its average using the weight function (\ref{weight}) and obtain
\begin{eqnarray}
A_{0}^2A_{1}{\rm e}^{i\left(2S_{0}+S_{1}\right)/\hbar}\int_{-c}^cdsdu\,w_{T_p}(s,u)\left({\rm e}^{i\Delta S_1/\hbar}-{\rm e}^{i\Delta S_2/\hbar}\right)=\mathcal{O}\left({\rm e}^{-\lambda T_0}\right).
\end{eqnarray}
The r.h.s.\ follows from Taylor expanding the exponential containing the second term in $\Delta S_1$ and is negligible for $T_0\sim T_1\sim T_H$.
This implies a cancellation of the corresponding contributions to the spectral determinant, that is,
all contributions from POs with one 2-encounter and multiple partial traversals along one of the shorter POs
involved.

\subsection{Orbits with at most two 2-encounters ($n\leq2$)}\label{2en}

Here we generalize the statements from the last subsection to POs with at most two encounters. We
include this intermediate step before addressing the general case of at most $n$ encounters 
(with $n$ integer usually of the order $1/\hbar$ for $T \sim T_H$, see Eq.~(\ref{napprox})),
because it is still much more transparent than the general case, and it provides additional insights 
compared to the above case $n \leq 1$.

\begin{figure}
\begin{center}
 \includegraphics[width=10cm]{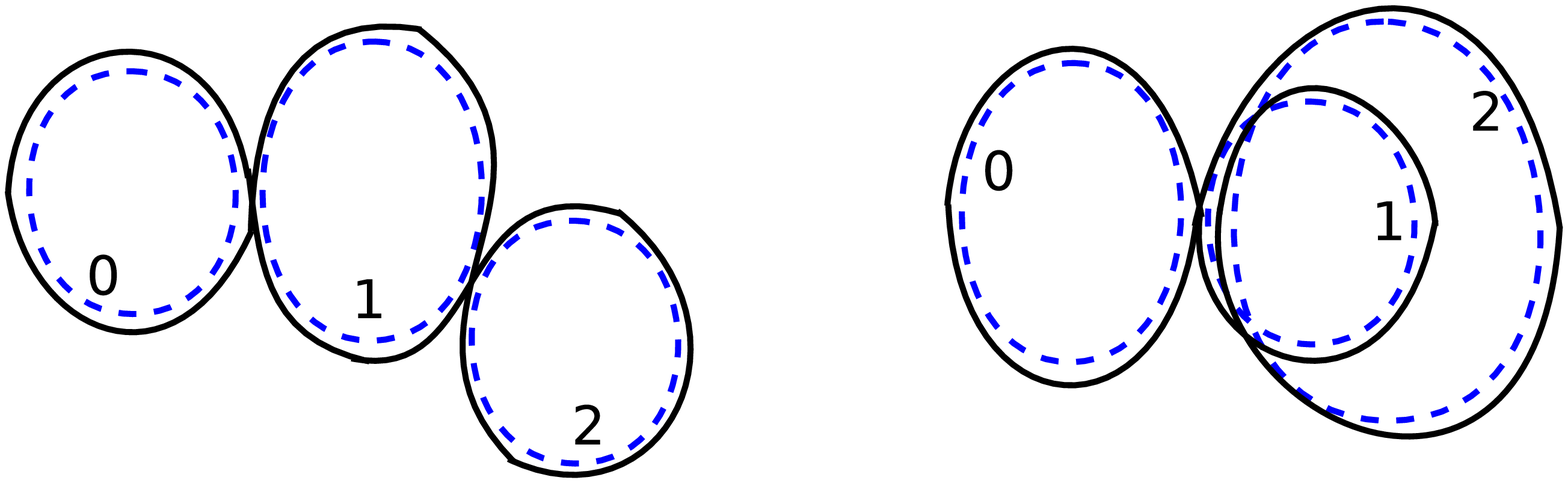}
 \caption{Long periodic orbit (solid curve) composed of three shorter periodic orbits ``0'', ``1'' and ``2'' (dashed)
by means of two separate (left) or overlapping (right) 2-encounters.}
\label{fig6}
 \end{center}
\end{figure}

We first note that we do not require the two 2-encounters to be separated from each other but also allow them 
to overlap, see right panel of Fig.\ \ref{fig6}, in contrast to the nomenclature from previous literature \cite{Mul04,Heu07} 
where this situation is referred to as a 3-encounter.
Again we first consider item 1 at the end of Subsec.~2.1. 
Using the same arguments as above given between Eqs.\ (\ref{overcon}) and (\ref{overcon1})
we obtain for the contribution (of PO configurations as in Fig.~\ref{fig6}) to the spectral determinant 
\begin{eqnarray}\label{con56}
&& A_{{012}}{\rm e}^{iS_{{012}}}-A_{{01}}A_{2}{\rm e}^{i\left(S_{{01}}+S_{2}\right)/\hbar}
\nonumber\\ &&
-A_{{12}}A_{0}{\rm e}^{i\left(S_{{12}}+S_{0}\right)/\hbar}
+A_{0}A_{1}A_{2}{\rm e}^{i\left(S_{0}+S_{1}+S_{2}\right)/\hbar} \nonumber\\ &&
\approx A_{0}A_{1}A_{2}{\rm e}^{i\left(S_{0}+S_{1}+S_{2}\right)/\hbar} \left[1+2\frac{T_{{012}}}{T_H}
+\left(\frac{T_{{012}}}{T_H}\right)^2\right].
\end{eqnarray}
In the equation above we get two contributions from pseudo orbits consisting of two orbits.
Note that in the case of large $n$ discussed below we obtain a sum  over the contributions 
from many different encounters for the same constituents. The fact that $n$ is eventually large justifies the averaging 
by $w_{01}(s,u)$ introduced above. Expression (\ref{con56}) 
 also holds true for overlapping encounters (Fig.\ \ref{fig6} right),
 as only the (still) connected and the split off orbit need to be distinguished.
Comparing the results above to the ones for the pseudo orbits formed by "0", "1", "2" on quantum graphs we again find 
direct correspondence to the ones in Eq. (\ref{con56}) for $T_{012}=T_H$; concerning the stability amplitudes remember Eq.\ (\ref{eq11}). 
%
%

Contrary to the last subsection (where only "0" and "1" were occuring and hence not allowing for resummation) 
here several pseudo orbits exist that possess durations between $T_H/2$ and $T_H$
consisting of three bonds  for the corresponding graphs. 
As shown in Subsec.\ \ref{graphs} the two contributions (from one fully connected PO and from a pseudo orbit with two components) 
can be directly added up rendering these contributions 
equal to the complementary ones from orbits consisting of one bond.

Turning to general chaotic systems we obtain by a calculation analogous to the one in Eq.\ (\ref{con56}) powers 
of ratios of $T_{{012}}/T_H$ in that case. 
In chaotic systems where the encounters are distributed in an ergodic manner such a ratio can be interpreted 
as the probability for an orbit "012" (denoted here by $p$) of duration $T_p$ to reach a certain encounter 
(instead of $T_H$ where the encounter is reached for sure). 
We relate this contribution from pseudo orbits of duration $T_p$ to the one from complementary
 pseudo orbits of duration $T_p'=T_H-T_p$. Their stability amplitudes are $2^{-(T_H-T_p)/(2\bar{T})}$ 
and $2^{-T_p/(2\bar{T})}$, respectively, with $\bar{T}$ the average temporal distance between encounters. 
The additional number of the pseudo orbits that can be formed in the latter case is given by 
$2^{(T_p-T_H/2)/\bar{T}}$. 
This number is obtained as follows: The duration difference between $(T_H-T_p)/2$ and $T_p/2$ is $T_p-T_H/2$. 
The number of encounters traversed during this time is $(T_p-T_H/2)/\bar{T}$. At each encounter
the trajectory branches associated with two possibilities, in total $2^{(T_p-T_H/2)/\bar{T}}$ possibilities. 
Multiplying this number by the stability amplitude $2^{-T_p/(2\bar{T})}$ of an orbit with duration $T_p$ 
we obtain the amplitude of an orbit of duration $(T_H-T_p)/2$. 
Taking into account that the actions of the pseudo orbits of durations $T_p$ and $T_p'$ sum up to the 
action of an orbit of duration $T_H$ we see in view of relation (\ref{nav})
that these two contributions to the spectral determinant are indeed complex conjugate to each other.

Orbits involving multiple traversals of shorter orbits again do not contribute to the spectral determinant 
due to the same arguments as given in the preceding subsection.

\subsection{Orbits with at most $n$ 2-encounters}
\label{nen}

\begin{figure}
\begin{center}
\includegraphics[width=9cm]{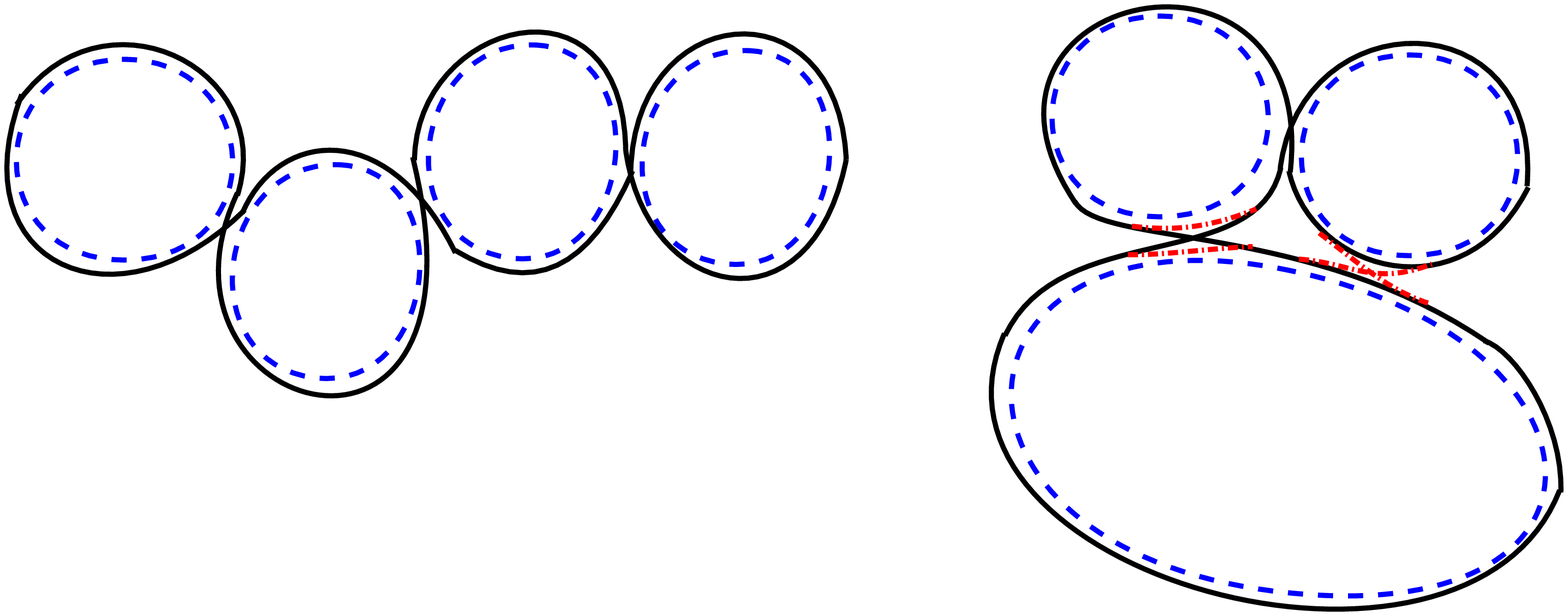}
\end{center}
\caption{
Left panel: One long, fully connected periodic orbit (full line) surrounds all shorter periodic orbits 
(dashed lines). Right panel: Two different configurations of connected orbits 
can be constructed surrounding the shorter periodic orbits. (indicated by the full and dashed dotted lines).}
\label{fig10}
\end{figure}

Here we consider resummation for the general case of a system with at most $n$ 2-encounters.
We analyse exemplarily the case of one fully connected orbit without repetitions, 
see the left diagram in Fig.\ \ref{fig10}.  We will comment on how to generalize this analysis
also to cases where several connected orbits exist, {\em e.g.} the right diagram in Fig.\ \ref{fig10}, 
at the end of this subsection.

For the case of a fully connected PO with at most $n$ two 2-encounters we can proceed in an analogous manner as above. 
At first, we have one fully connected orbit obtained by combining the pseudo orbits at all encounters. Furthermore,  
consider pseudo orbits composed of two components where at all encounters except for one the (shorter) pseudo 
orbits are combined, we have $n$ possibilities to choose the encounters where the pseudo orbit is already combined. 
This factor $n$ replaces the factor 2 obtained in Eq.\ (\ref{con56}) in Subsec.~\ref{2en}. 
Considering pseudo orbits consisting of three components we have $n(n-1)/2$ possibilities 
to choose the encounters where the orbit is not split to obtain the pseudo orbit:
$n$ for choosing the first encounter with combination of pseudo orbits and $n-1$ for choosing the second.
The factor $1/2$ enters because they cannot be distinguished. This procedure can be carried on for 
pseudo orbits consisting of more components.
As shown for graphs in Ref.~\cite{Wal13} we thus get a coherent sum of contributions from pseudo orbits
containing no partial repetitions. Their overall number is
\begin{equation}
\label{eq12}
\sum_{\nu=0}^n\left(\begin{array}{c}n \\ \nu\end{array}\right)=2^n \, ,
\end{equation}
where the contributions explained in Eq.\ (\ref{con56}) correspond to $n =2$ in the equation above.
Using Eq.\ (\ref{eq11}), the overall factor $2^n$ in Eq.~(\ref{eq12}) again cancels the stability 
amplitude $2^{-n}$ resulting from the $2n$ links between the encounters in this case. This leads again 
to two complementary contributions to the spectral determinant complex conjugated to each other yielding
together a  real result.

Here we assumed that 
the exponential increase in the number of POs of duration $T$ proportional to ${\rm e}^{\lambda T}$
is caused by branching of the POs at the $n$ encounters (with action differences of order $\hbar$). 
Such an assumption (see also Eq.~(\ref{eq11})) is safely fulfilled for long orbits of durations 
of order $T_H$ as they cover the energy shell on scales of order $\hbar$ leading to encounters with 
action differences that small that they contribute to Eqs.\ (\ref{suint},\ref{siint}).

As mentioned at the beginning of this subsection this analysis was done for graphs with only one 
fully connected orbit traversing the full graph without repetitions.
As we know first from our calculations for graphs \cite{Wal13} that the resummation can be done 
for several fully connected orbits without repetitions covering the full graph in a similar way, 
provided we obtain a minus sign when splitting the orbit at a 2-encounter and second that we 
have in the case of continuous dynamics the same ingredient; {\em i.e.}\ a minus sign resulting 
from splitting the orbit at every 2-encounter; 
the resummation can be performed in the same way for continuous dynamics as for graphs.

Concerning the resummation of pseudo orbits shorter than the $T_H$ we can proceed in the 
very same way as described in the last subsection. The same also applies to the cancellation of
the contribution from POs with partial repetitions.

Throughout the above derivations we showed the analogy between the resummation procedures for 
quantum graphs on the one side and for systems with chaotic dynamics on the other side. 
These arguments hold for arbitrary values of $n$ and thus also for values consistent with 
Eq.\ (\ref{napprox}).


\section{Implications for quantum chaotic many-body dynamics}
\label{MB}
So far we implicitly assumed chaotic dynamics in the low-dimensional phase space of single-particle 
systems. Recently, interacting many-body (MB) quantum systems with a chaotic classical limit have
gained considerable interest across different communities in theoretical physics. 
The semiclassical methods forming the basis of our present work are not limited to single-particle 
dynamics: In the MB context a semiclassical version of the Gutzwiller-van Vleck propoagator 
was derived for bosonic \cite{EnglI} and fermionic \cite{EnglII} MB systems, as well as spin 
systems \cite{Waltner}. Specifically, periodic-orbit expansions could be deduced for the 
density of states of interacting bosonic quantum fields on a lattice \cite{EnglIII,Dubertrand}. 
There the semiclassical regime corresponds to the limit of large particle numbers $N$ where $1/N$ takes
the role of an effective $\hbar$. In the classical (mean-field) limit, reached for $N\rightarrow \infty$, 
these systems are described by non-linear wave equations. Assuming  this  mean-field dynamics to
be fully chaotic, in Refs.~\cite{EnglIII,Dubertrand} it was shown that the fluctuating part of the 
bosonic MB density of states can then be represented in terms of a semiclassical 
trace formula that has the very same structure as Gutzwiller's trace formula \cite{Gut90} 
for chaotic single-particle quantum systems. 
Periodic mean field solutions of the non-linear wave equations take the role of the classical single-particle 
POs in Gutzwiller's trace formula, including corresponding stabilities and actions. This close formal
analogy has then allowed for rather straight-forwardly generalizing the semiclassical
calculation of the spectral MB form factor \cite{EnglIII,Dubertrand}, more generally, providing a dynamical 
explanation of universal spectral statistics in chaotic MB systems 
\cite{foot}.
To this end the real-valued spectral determinant was again assumed and employed \cite{Dubertrand}.

Due to this formal analogy between the semiclassical treatment of single-particle and
MB dynamics, our arguments and derivations, put forward in Sec.~\ref{resumgen},  providing a semiclassical interpretation of 
the real-valuedness of the spectral determinant can be taken over to the MB case.
The main underlying aspect involves a hierachical structure of POs, intertwined through encounters
that provide the mechanism for long POs to be composed of (mulitple repetitions along) shorter POs.
In the MB context coresponding encounters exist between (periodic) mean-field modes in high-dimensional MB phase space
and provide the underlying semiclassical mechanism for true quantum MB interference. In other words, 
in large-$N$ quantum systems entanglement is created (at time scales beyond the Ehrenfest time $\sim \log N$)
between different mean field solutions as they are linked to each other through encounters \cite{Rammensee}.
It has been recently shown \cite{Tomsovic} that corresponding semiclassical techniques 
indeed quantitatively describe MB quantum interference at such post-Ehrenfest time scales.

Under the assumption of hyperbolic mean-field dynamics, this altogether indicates to re-interprete 
items 1.\ and 2.\ in Subsec.~\ref{subsec22}: 

1.\ Contributions to the MB spectral determinant from pseudo 
orbits (based on periodic mean-field solutions) of duration $T< T_H$ are complex conjugate to contributions 
from pseudo orbits of length $T_H-T$ covering the complementary phase space area. 

2.\ All contributions to the MB spectral determinant from pseudo orbits involving multiple partial traversals
mutually cancel. In particular, this implies that pseudo orbits with durations $T > T_H$ that necessarily involve
such multiple traversals do not contribute. 


\section{Conclusions}
\label{conclu}

In this paper we have demonstrated how to generalize a resummation procedure for pseudo orbits in
quantum graphs to general quantum mechanical single-particle systems with
chaotic classical counterpart, with further implications to chaotic many-body systems.
In the semiclassical limit long orbits with durations of the order of or larger than the Heisenberg time 
fill the phase space such densely that the previously considered individual encounters loose their significance and 
are replaced by periodic orbits containing partial repetitions from surrounded shorter periodic orbits.
Thereby we demonstrate a corresponding Riemann Siegel relation for the spectral
determinant exploiting subtle classical correlation mechanisms of periodic pseudo orbits. 
Using this relation and following the calculation in Ref.\ \cite{Kea07}
we can further justify the validity of the Bohigas Giannoni Schmit conjecture based on purely dynamical grounds.
Establishing a dynamical mechanism for a real-valued spectral determinant moreover implies a better
understanding of the spectral form factor and its saturation in the late time limit beyond the Heisenberg time,
based on purely semiclassical arguments.

The analogy to quantum graphs shown in this paper was established for chaotic systems without 
time-reversal symmetry and graphs with directed bonds. However, it can be generalized in a 
straightforward manner to systems with time-reversal symmetry. In that case, apart from the 
pseudo-orbit pairs existing without time-reversal symmetry, there exist additional orbit pairs 
traversed in opposite directions. Their contributions add up and cancel in the same way as explained
in the main part of the paper. Additionally, two further possibilities arise, as the short multiply traversed
orbits, {\em e.g.}\ ``0'' and ``1'' in Fig.\ \ref{fig2}, can traverse the encounter region {\em 
relative to each other} in two different directions. However, for these configurations we can identify further 
encounters in the same orbit where the direction of traversal relative to each other is not switched and then 
add up or cancel their contributions as described in the main part. 

Altogether the resummation procedure in the presence of time-reversal symmetry can be obtained as a straightforward 
extension of the one for systems without time-reversal symmetry. The same applies also to systems with weak spin-orbit coupling 
(see Ref.~\cite{Bolte} for a semiclassical treatment) as in that case the classical dynamics 
reamins unchanged.

The Riemann Siegel relation  reflects unitarity and holds true independent of the character of the dynamics 
of the underlying classical system. However the construction of the diagrams considered here assumes exponentially 
approaching and deviating trajectories as long as the dynamics is linearizable and ergodicity holds for the 
underlying system. It would be thus interesting to identify the relevant correlations in the case of
integrable and mixed dynamics.

Moreover, as outlined in Sec.~\ref{MB} semiclassical techniques as the one used provide
powerful tools to treat quantum many-body systems with complex classical many-particle dynamics in
the wide terrain of large particle number or small $\hbar$. 
Hence it is of interest to further make use of the benefits resulting from semiclassical resummation of the 
spectral determinant for the calculation of quantum spectra of many-particle systems with focus on
effects typical for them such as indistinguishability of the particles or collective 
many-body dynamics \cite{Akila}. More generally, to better understand spectral features 
of the large variety of many-body dynamics remains as a future challenge.


\section{Acknowledgements}
We thank Petr Braun and Gregor Tanner for numerous discussions and furthermore Petr Braun for many valuable 
suggestions to improve the presentation of this manuscript. We further acknowledge support by the 
{\em Deutsche Forschungsgemeinschaft} through projects Gu431/9-1 and Ri681/14-1 (within the ''Dreiburg cooperation'').


\section*{Appendix: Estimate for the number of encounters}
\appendix
We give an estimate for the number $n$ of encounters (see Eq.\ (\ref{napprox})) based on arguments given in 
Ref.\ \cite{Sie01}. There the number of self-crossings $dn$ of an orbit of duration $T$ 
under an angle in the range $\left[\epsilon,\epsilon+d\epsilon\right]$ was estimated for a billiard of 
area $A$ and $\epsilon\ll1$ as
\begin{equation}\label{numb}
dn\sim\frac{T^2v^2}{4\pi A}d\epsilon^2
\end{equation}
with the magnitude of the velocity $v$. Employing the expression for the action difference 
\begin{equation}\label{numb1}
\Delta S=\frac{p^2\epsilon^2}{2m\lambda}
\end{equation}
with mass $m$ and Lyapunov exponent $\lambda$ between orbit and partner orbit in Fig.~\ref{fig2} the 
expression (\ref{numb}) can be rewritten as
\begin{equation}
dn\sim\frac{\lambda T^2}{2\pi mA}d(\Delta S) \, .
\end{equation}
Taking into account only action differences up to order $\hbar$ we obtain by integration 
for the number of encounters of a PO at $T=T_H$
\begin{equation}
n\sim\frac{\lambda T_H^2\hbar}{2\pi mA}=\frac{\lambda T_H}{2\pi} = 
\frac{m\lambda A}{2\pi \hbar} \, ,
\end{equation}
providing Eq.\ (\ref{napprox}) of the main text.

\end{document}